# Advances in the experimental exploration of water's phase diagram

Christoph G. Salzmann*

Department of Chemistry, University College London, 20 Gordon Street, London WC1H 0AJ, United Kingdom; E-mail: c.salzmann@ucl.ac.uk

**Abstract**

Water's phase diagram displays enormous complexity with currently 17 experimentally-confirmed polymorphs of ice and several more predicted computationally. For almost 120 years, it has been a stomping ground for scientific discovery and ice research has often been a trailblazer for investigations into a wide range of materials-related phenomena. Here, the experimental progress of the last couple of years is reviewed, and open questions as well as future challenges are discussed. The specific topics include the polytypism and stacking disorder of ice I, the mechanism of the pressure amorphization of ice I, the emptying of gas-filled clathrate hydrates to give new low-density ice polymorphs, the effects of acid / base doping on hydrogen-ordering phase transitions as well as the formation of solids solutions between salts and the ice polymorphs, and the effect this has on the appearance of the phase diagram. In addition to continuing efforts to push the boundaries in terms of the extremes of pressure and temperature, the exploration of the 'chemical' dimensions of ice research appears to now be a newly emerging trend. It is without question that ice research has entered a very exciting era.

## I. Introduction

From the poles of the inner planets and comets to the moons of the gas giants and distant dwarf planets, ice is a constant companion across the entire Solar System.[1, 2] On Earth, the freezing of water to give the 'ordinary' ice I*h* is one of the most frequently observed phase transitions and the experience of snowfall is an unforgettable natural spectacle. However, the formation of ice can also wreak havoc in aviation and road traffic, interrupt power lines and destroy infrastructure. Hence, tremendous research efforts have recently been directed towards developing anti-icing surfaces.[3, 4]

From the scientific perspective, ice is a highly complex material that has been a trailblazer for research into a wide range of materials-related phenomena. These include, for example, pressure-induced amorphization,[5] negative thermal expansion,[6] formation of a large variety of inclusion compounds (*i.e.* clathrate hydrates)[7, 8] and a remarkably rich phase diagram.[9] A more detailed understanding of the suppression of ice formation in biological samples has led to the successes of cyro-electron microscopy, for which the 2017 Nobel Prize in Chemistry has been awarded.[10] The tremendous versatility of ice as a material is illustrated by its recent use as a solid support in chromatography.[11]

A defining structural aspect of ice I*h* is that it is actually not a truly crystalline material. In fact, its tetrahedrally hydrogen-bonded water molecules display orientational disorder, which gives rise to a molar configurational entropy estimated as $R$ ln (3/2) by Pauling.[12] Due to the disorder with respect to the hydrogen atoms, ice I*h* is classified as a hydrogen-disordered phase of ice.

The first high-pressure phases of ice were discovered by Tammann and Bridgman in the early 20$^{th}$ century.[13, 14] The convention in the field is that a newly discovered ice polymorph is labelled with the next available Roman numeral. The evidence for a new phase of ice has to be based on crystallographic or at least spectroscopic data.[15] The two most recently discovered polymorphs of ice are ices XVI and XVII, which were obtained by removing gaseous guest species from clathrate hydrates.[16-18] Ice VII has very recently been identified as inclusions in diamond and it is now, together with ice I*h*, officially recognized as a mineral.[19]

Figure 1 shows the up-to-date phase diagram of ice in the pressure range up to 100 GPa. It can be seen that ice I*h* is not the only hydrogen-disordered phase of ice. In fact, all polymorphs that can be crystallized from the liquid are hydrogen-disordered including the two metastable phases IV and XII. Upon cooling the hydrogen-disordered ices, phase transitions to the corresponding hydrogen-ordered counterparts can be observed, which display orientational order of the water molecules. If complete hydrogen order can be established, then the configurational entropy of such a phase is equal to zero.[9] However, hydrogen ordering ice is typically a difficult process since it relies on highly cooperative reorientation dynamics of the water molecules. In fact, only ices III and VII undergo spontaneous hydrogen ordering upon cooling whereas the formation of ices XI, XIII, XIV and XV requires doping with bases or acids.[20-22] The topologies of the hydrogen-bonded networks remain unchanged upon hydrogen ordering, which means that there are in principle pairs consisting of a hydrogen-disordered and a hydrogen-ordered phase for a given network topology.[9] However, for stacking-disordered ice I (ice I*sd*) as well as for ices XVI and XVII, the hydrogen-ordered counterparts have so far not been identified. In the case of ice IV, it has been shown that doping with hydrochloric acid leads to the appearance of a weak endothermic feature in calorimetry upon heating, which could indicate a phase transition from weakly hydrogen-ordered ice IV to hydrogen-disordered ice IV.[9] Ice II is unique in the sense that it is the only hydrogen-ordered phase of ice that does not have a hydrogen-disordered counterpart.[23] Remarkably, it has been predicted computationally that the phase boundary between ice II and its hydrogen-disordered counterpart, ice II*d*, is located in a region of the phase diagram where the liquid water is stable (*cf.* Figure 1).[24] Upon compression of ice VII, the hydrogen bonds become symmetric in ice X and the phenomenon of hydrogen order / disorder is no longer relevant since the water molecules lose their molecular character.[25, 26]

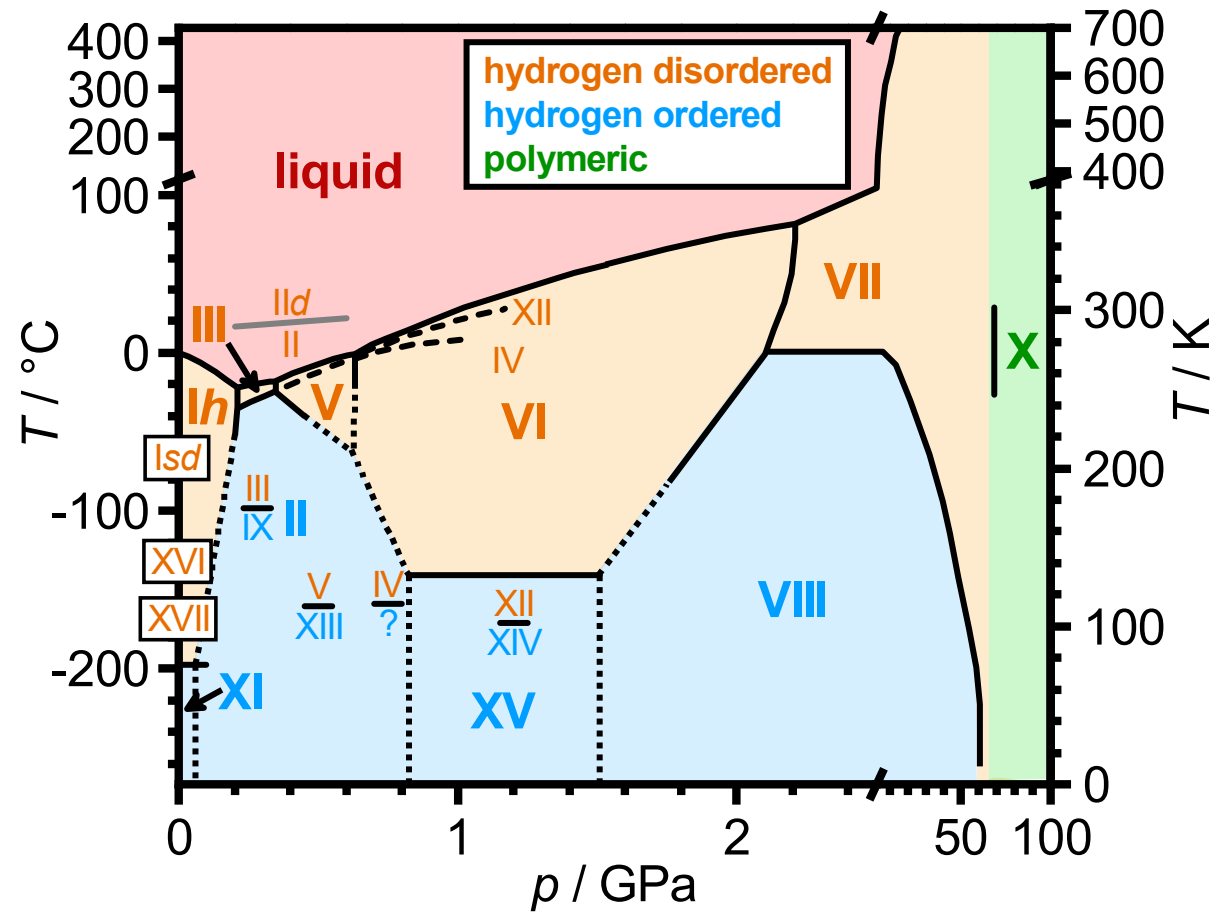

***Fig. 1*** *Phase diagram of ice including the regions of thermodynamic stability of liquid water (red), the hydrogen-disordered (orange) and hydrogen-ordered (blue) phases of ice as well as the 'polymeric' ice X (green). Stable phases of ice are highlighted by large bold Roman numerals whereas metastable states are indicated by a smaller font size. Black dotted lines are extrapolated phase boundaries at lower temperatures. Dashed lines indicate the metastable melting lines of ices IV and XII. The phase boundary between ice II and IId (gray) has been predicted computationally.[24] Adapted from refs [9, 15-18, 24].*

Figure 2 showcases the enormous structural diversity of the 'ice family'. Ices I*h*/XI, II and XVII can be classified as 'open-channel' structures. In case of ice I*h*/XI, the channels are next to each other whereas in ice II, individual 'ice nanotubes' are hydrogen-bonded to one another. The ice XVII channels have a spiral geometry, which reflects the chirality of the crystal structure.[17, 18] The ice III/IX network contains 4-fold spirals, which means that this structure is chiral as well since the spirals can be either left- or right-handed. The hallmark structural feature of the ice IV network are flat six-membered rings that are interpenetrated by a hydrogen bond. In general, such interpenetrating structural features enable denser network structures to be realized. The ice V/XIII network is the structurally most complex with several of the hydrogen-bonded chains running along the crystallographic axes and ring sizes ranging from four to twelve.[9] Interpenetration of two individual networks are found in ice VI/XV and ice

VII/VIII/X. The ice VII/VIII/X network is the densest ice structure currently known with a first-neighbor coordination number of eight including four hydrogen-bonded and four non-hydrogen-bonded oxygen atoms. The ice XII/XIV network is the densest in absence of interpenetrating elements.[27] In the projection shown in Figure 2, the network topology resembles the Cairo tiling. Yet, the 5-membered rings of the Cairo tiling are in fact seven-membered rings in ice XII/XIV due to the highlighted 'zig-zag' chains. Ice XVI has the structure of the empty cubic structure II clathrate hydrate with two different types of cages: small $5^{12}$ cages in the form of pentagonal dodecahedra and large $5^{12}6^4$ cages, which contain four additional six-membered rings.[16]

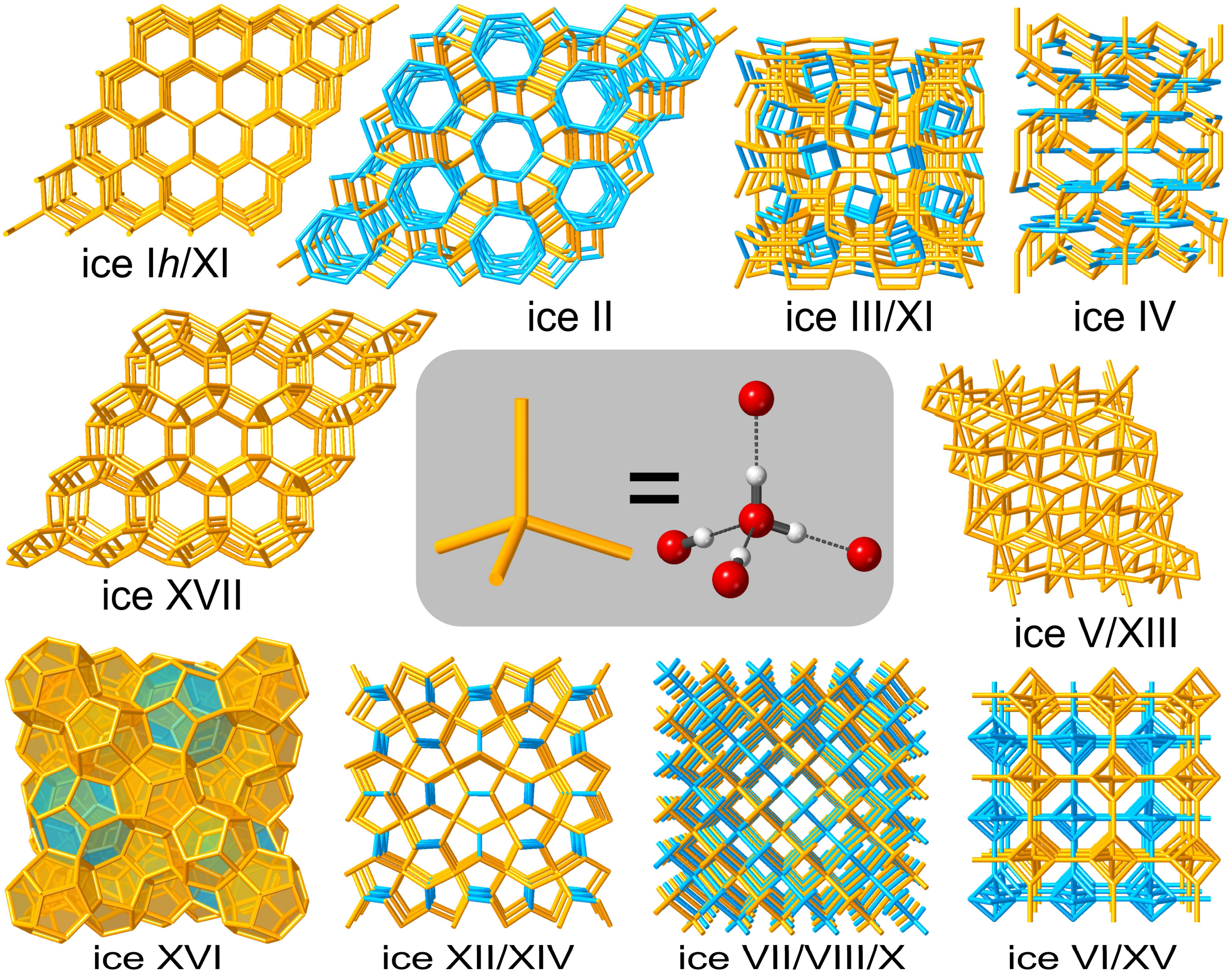

***Fig. 2*** *The hydrogen-bonded networks of the various polymorphs of ice. The inset shows how the structure of a local hydrogen-bonded environment of a water molecule corresponds to the orange stick-representation used in this figure. Some of the network topologies are shared between the hydrogen-disordered phases and their hydrogen-ordered counterparts. Noteworthy structural features are highlighted in cyan. These include the six-membered rings of the 'ice nanotubes' in ice II, the 4-fold spirals in ice III/IX, the interpenetrated six-membered rings in ice IV, the second independent hydrogen-bonded networks in ice VI/XV and ice VII/VIII/X, and the 'zig-zag' chains of ice XII/XIV. The* $5^{12}$ *and* $5^{12}6^4$ *cages of ice XVI are shown as orange and cyan polyhedra, respectively. Adapted from refs [9, 16, 17].*

Despite the remarkable structural complexity displayed in Figure 2, it seems likely that we are not yet looking at the complete picture and that further ice polymorphs will be discovered in the future. Most recently, 74963 potential ice structures were identified computationally on the basis of a zeolite database.[28] Particular progress has also recently been made with the computational prediction of low- and even ultra-low-density 'aeroice' network topologies.[29-31] How and if these structures can be prepared experimentally is of course entirely unclear at present.

This perspective article reviews the recent experimental advances in exploration of the polymorphism of ice and future directions of research will be discussed. Specifically, the focus will be placed on pure ice first and the topics of stacking disorder in ice I and the mechanism of the pressure-induced amorphization of ice I will be introduced. Then, newly emerging 'chemical' directions in ice research will be explored such as the emptying of clathrate hydrates, acid / base doping-induced hydrogen ordering as well as the formation of solid solutions between the ice polymorphs and ionic species.

**II. Stacking disorder and polytypism of ice I**

Kuhs *et al.* realized on the basis of diffraction data that so-called cubic ice I samples were in fact not entirely cubic but contained hexagonal stacking as well.[32] The structure of such a stacking-disordered ice I (ice I*sd*) is shown in Figure 3. Ice I*sd* contains the same layers of hydrogen-bonded water molecules as found in ice I*h*. However, in addition to the hexagonal stacking of these layers as found in ice I*h*, ice I*sd* also contains cubic stacking sequences. The two geometric recipes for stacking result in different conformations of the six-membered rings linking two adjacent layers as shown in Figure 3. Specifically, hexagonal stacking results in six-membered rings with the 'boat' conformation whereas the 'armchair' conformation is found for cubic stacking. From the perspective of the ice I*h* channels shown in Figure 2, cubic stacking leads to termination of the channels.

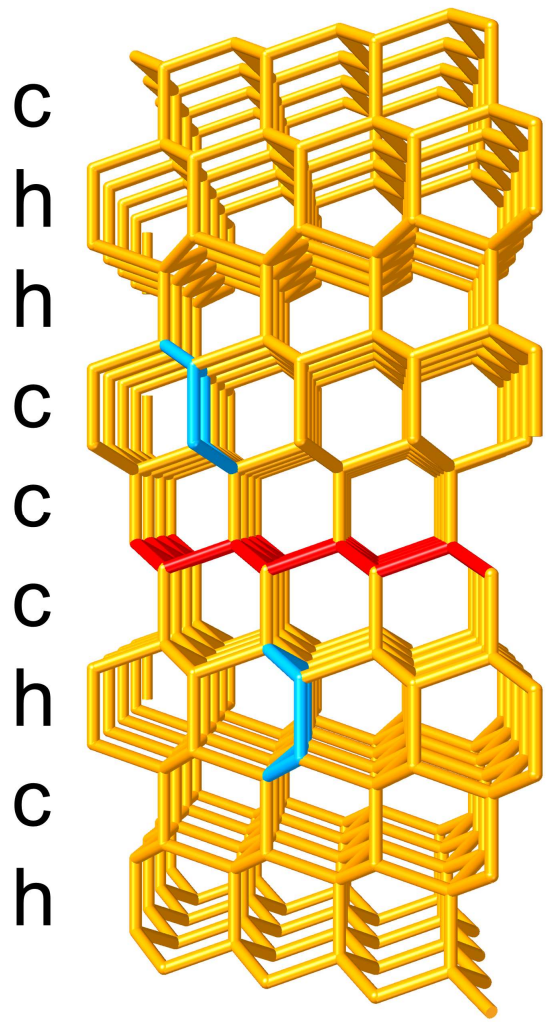

***Fig. 3*** *Hydrogen-bonded network of stacking disordered ice I (ice Isd). Hexagonal and cubic stacking are indicated by "h" and "c", respectively. One of the layers is shown in red. The different conformations of the six-membered rings resulting from the two types of stacking are highlighted in cyan.*

Work in this area has received a major boost in the last couple of years, which was mainly due to advances in modelling the diffuse diffraction features that arise from the stacking disorder.[33-36] Based on such analyses, it is possible to determine the 'cubicity', *i.e.* the percentage of cubic stacking present in an ice I*sd* sample. The percentage of hexagonal stacking or 'hexagonality' is then simply 100% minus the cubicity. In principle, the cubicity is free to take continuous values from 0 to 100% corresponding to structures from fully hexagonal to fully cubic ice I. Remarkably, depending on how ice I*sd* is prepared, vastly different cubicities are obtained.[33-36] Yet, at present it is unclear which factors determine the cubicity of the product ice I*sd* material. As far as bulk samples are concerned, the most cubic ice I*sd* with a cubicity of 73.3% has been obtained upon heating ice II at ambient pressure.[34] Freezing nanodroplets of supercooled water in a vacuum has resulted in a cubicity of 78%, which is at present the most cubic ice I*sd* identified so far.[37] In addition to determining the cubicity of an ice I sample, the analysis of diffraction data can also reveal information about memory or 'Reichweite' effects within the stacking sequences. So far, all characterized ice I*sd* samples displayed either random stacking or a tendency for staying with a given type of stacking within a sequence rather than preferential switching between cubic and hexagonal stacking.[33-36]

In addition to fitting X-ray or neutron diffraction data, it has been shown that stacking disorder in ice I can be detected from its Raman spectrum.[38] This is exciting since it could facilitate the remote detection of ice I*sd*, which may exist, for example, in the upper atmosphere of Earth or on comets. The occurrence of snowflakes with 3-fold symmetry has been discussed as potential evidence for the existence of ice I*sd* in the atmosphere.[39] It has also been shown that the cubicity of ice I*sd* can be determined from experimental pair-distribution functions.[40] Furthermore, the ice I*sd* to ice I*h* phase transition was recently followed with dielectric and NMR spectroscopy,[41] and the advantages of using Markov chains for the description of stacking disorder have been presented.[42]

At present, it is not clear if fully cubic ice I (ice I*c*) can be prepared. However, current efforts are certainly directed towards achieving this goal. As recently illustrated in the case of the isostructural silicon,[43] the use of a suitable template, on which the ice I*c* can be grown, could be a promising way forward. Moving beyond ice I*c*, it is intriguing to speculate if ice I polytypes with higher-order memory effects can be prepared such as the 4H, 6H and 9R polytypes according to the Ramsdell notation. In case of silicon, the 4H polytype, in which cubic and hexagonal stacking strictly alternates, was recently prepared using a high-pressure procedure.[44] Potentially, the ice I family of polytypes could therefore be much larger than currently appreciated.

**III. Mechanism of the pressure amorphization of ice I**

Seminal work by Mishima *et al.* has shown that ice I*h* undergoes pressure-induced amorphization upon compression at liquid-nitrogen temperature (77 K).[5] It has since been an open question if the resulting high-density amorphous ice (HDA) represents a glassy state of high-pressure water or if it should be seen as a collapsed crystalline state.

Recently, Shephard *et al.* compressed ammonium fluoride ($NH_4F$) I*h* and I*sd* samples at 77 K.[45] $NH_4F$-I is isostructural with ice I and contains hydrogen bonds of very similar strength. The $NH_4F$ samples displayed very similar pressure collapses in terms of volume changes and transition pressures compared to ice I. However, the resulting $NH_4F$ materials were not amorphous but crystalline $NH_4F$-II, a high-pressure phase of $NH_4F$ isostructural with ice IV. A very elegant mechanism exists for the transition from ice I-type networks to the ice IV topology, which has been called the Engelhardt-Kamb collapse in ref. [45]. During this phase change, the hydrogen bonding within the layers of ice I remains unchanged (*cf.* Figure 3). However, the hydrogen bonds to the next layers above and below are broken, and reformed with water molecules two layers above and below the original layer, which is how the interpenetration of the six-membered rings of ice IV is achieved (*cf.* Figure 2). Overall, only a quarter of the hydrogen bonds need to be broken during this collapse, which achieves a 37% increase in density. However, because of the hydrogen disorder in ice I, the chances of successfully breaking a hydrogen bond and reforming it are only 50%. Unlike $NH_4F$, ice I does therefore not follow the Engelhardt-Kamb collapse through entirely and the resulting HDA was classified as a 'derailed' state along the ice I to ice IV pathway.[45]

Upon heating HDA around its 'natural' pressure of about 1 GPa, the crystallization to ice IV has the smallest activation energy,[46] which illustrates that with the help of unfrozen reorientation dynamics at higher temperatures, the original 'derailment' due to the hydrogen-bond mismatches can be brought back on track and the sample finally transforms to ice IV.[45] According to this scenario, the reason why ice I undergoes its well-known pressure-induced amorphization at low temperatures is its hydrogen-disordered nature combined with the very slow reorientation dynamics at 77 K. Using nudged-elastic-band DFT calculations, it was shown that the hydrogen-ordered ice XI could indeed transform to ice IV upon low temperature compression.[45]

Lin *et al.* recently showed with X-ray diffraction that a crystalline pre-state exists *prior* to the onset of pressure amorphization upon compression at 100 K.[47] The structure of the pre-state was rationalized in terms of ice I*h* that has experienced shearing of its layers within the basal plane in line with softening of the $C_{66}$ elastic modulus. This suggestion is consistent with the Engelhardt-Kamb collapse since the interpenetration of the six-membered rings requires the layers in ice I*h* to shift (*cf.* Figure 2). Interestingly, the shifting of layers is not required if cubic stacking is present, and in line with this, ice I*sd* shows a slightly lower onset-pressure for pressure amorphization.[45] Lin *et al*. conclude that HDA is 'an intermediate state in the phase transition from the connected H-bond water network in low pressure ices to the independent and interpenetrating H-bond network of high-pressure ices'.[47] Most recently, it has been shown that ice I*sd* can be pressure-amorphized at temperatures up to 174 K if compression rates close 50 GPa $s^{-1}$ are used.[48] This allows the phase transitions to ices II and VI to be suppressed, which nicely illustrates the kinetic origin of the pressure amorphization of ice I.

On the basis of the 'derailed' nature of HDA, the previously suspected 'crystalline remnants' in HDA[49] can now be explained in terms of local regions where due to maximal mismatches for breaking and reforming hydrogen bonds a quite ice I-like local structure is retained. Similarities of the local structures of HDA and ice IV were also found in a recent analysis of the experimental pair-distribution functions[50] as well as in a molecular dynamics study of the pressure-amorphization process.[51] Overall, the evidence seems to mount that HDA, as prepared by compression of ice I at 77 K, is a collapsed crystalline material that does not have a thermodynamic connection with the liquid under pressure. However, if annealing HDA under pressure causes this to change is an open question. A particular current focus in this respect is placed on expanded HDA (eHDA) first identified by Nelmes *et al*.[52, 53]

**IV. New ice polymorphs by emptying clathrate hydrates**

An exciting avenue for preparing new ice polymorphs has recently been presented by Falenty *et al*.[16] Ice XVI was prepared by removing the neon guest from neon-filled cubic structure II clathrate hydrate. The complete removal was achieved by continuous pumping for several days on small particles of the filled clathrate hydrate at 142 K, which just a few degrees below the decomposition temperature to ice I*sd*. Compared to the neon-filled clathrate hydrate, ice XVI is less dense and it exhibits negative thermal expansion below 50 K. Most recently, it has been demonstrated that ice XVI can be filled with helium, which can also be removed upon heating in vacuum.[54] In principle, it may also be possible to remove guest species from other clathrate hydrates such as the cubic structure I or the hexagonal structure I clathrate hydrates. However, in order to form these, guest species larger than neon are typically needed,[8] which may be difficult to remove by pumping.

Del Rosso *et al.* achieved the emptying of the $C_0$ dihydrogen clathrate hydrate under vacuum below 120 K.[18] In contrast to the cages of ice XVI, the newly discovered chiral ice XVII has open channels (*cf.* Figure 2),[17, 55] which makes the removal of the gaseous guest species easier. As monitored with Raman spectroscopy, the dihydrogen was completely removed after pumping for 1-2 hours at ~120 K.[18] The removal of dihydrogen is reversible and there is also some evidence that the channels of ice XVII can be filled with dinitrogen as the samples are stored in liquid nitrogen. The dynamics of the dihydrogen molecules confined within the $C_0$ clathrate hydrate have been followed with inelastic neutron scattering.[56] Interestingly, the $C_0$ clathrate hydrate structure was recently found for $CO_2$ as well.[57] Yet, the removal of $CO_2$ by pumping to obtain ice XVII may prove difficult.

The most promising guest species in clathrate hydrates for subsequent removal by pumping seem to be dihydrogen, helium and neon. Both dihydrogen as well as helium form high-pressure clathrate hydrates corresponding to filled ice II structures.[7] Emptying these would therefore not lead to a new ice polymorph. However, in case of dihydrogen, it has been shown that a clathrate hydrate corresponding to a filled ice I*c* exists at pressures above 2.3 GPa.[58] Potentially, emptying this clathrate hydrate could enable the preparation of the elusive fully cubic ice I*c*.[59] Another approach for making new ice polymorphs could be the co-deposition of water and noble gases at low temperatures followed by the careful removal of the gas matrix upon heating.[60] As mentioned earlier, a range of ultra-low density ices and hence possible target polymorphs have recently been described computationally.[29-31]

## V. Acid / base doping of ice and its effect on hydrogen ordering

The various pathways that a hydrogen-disordered phase of ice can take upon cooling are schematically shown in Figure 4 in terms of enthalpy. At high temperatures, dynamic reorientation processes take place within the paraelectric hydrogen-disordered ices. As observed for ices I*h*, IV, V, VI and XII, the reorientation dynamics slow down upon cooling in such a fashion so that the ices 'fall out' of equilibrium at $T_g$(pure) and orientational glasses are obtained along pathway (1). The pure ices IV, V, VI and XII all show such glass transitions with heat capacity increases around 1 J $K^{-1}$ $mol^{-1}$.[9] In case of ice VI, a calorimetric study comparing $H_2O$, $D_2O$ and $H_2{}^{18}O$ samples has demonstrated that changes in reorientation dynamics indeed govern this type of glass transition.[61]

Salzmann *et al.* found that doping ices V and XII with hydrochloric acid (HCl) speeds-up the reorientation dynamics by introducing extrinsic $H_3O^+$ point defects so that once the hydrogen ordering temperatures ($T_{ordering}$) were reached upon cooling, the corresponding hydrogen-ordered ices XIII and XIV were discovered.[22] Slow-cooling DCl-doped $D_2O$ ice V at ambient pressure led to essentially fully hydrogen ordered ice XIII, corresponding to pathway (4), whereas in case of DCl-doped $D_2O$ ice XII, only a partially ordered ice XIV was obtained (pathway (3)).[22]

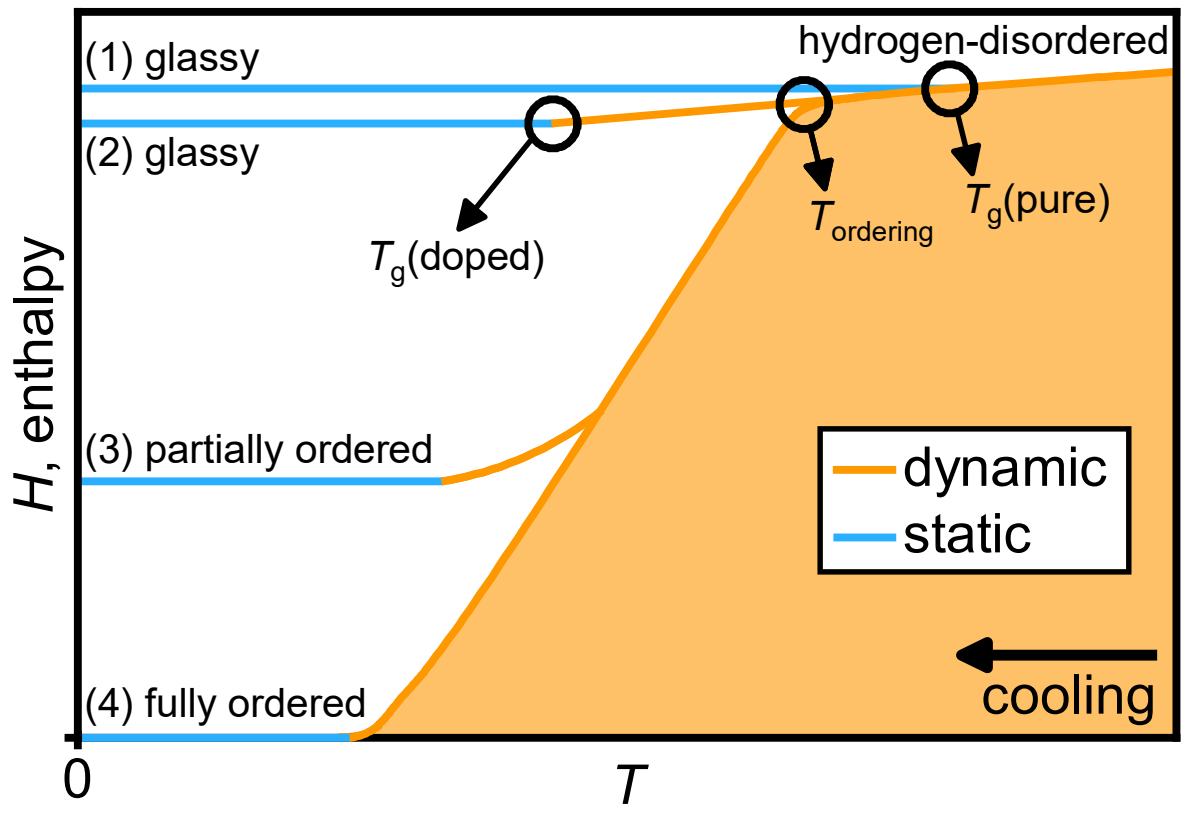

***Fig. 4*** *Schematic illustration of the various processes that can take place upon cooling pure or doped hydrogen-disordered ice leading to glassy, partially hydrogen-ordered or fully hydrogen-ordered ice. Adapted from refs [9, 62].*

On the basis of calorimetry, Köster *et al.* claimed that cooling HCl-doped $H_2O$ ice XII at 0.8 GPa leads to a 'complete loss of Pauling entropy' implying a transition from fully hydrogen-disordered ice XII to fully hydrogen-ordered ice XIV.[63] However, this claim was recently retracted and a mistake with the integration of the calorimetric data was conceded.[64] After reanalysis of the original calorimetry data, only a loss of 60% of the Pauling entropy was found.[64] Depending on how temperature is considered during the integration, the calorimetric change may in fact be as low as 51% of the Pauling entropy.[65]

The ice V to ice XIII was found to go along with a 66% loss of Pauling entropy.[65, 66] Since ice XIII is very highly ordered, this value reflects the partial hydrogen order already present within ice

V.[67] Most recently, a wide range of different acid and base dopants were tested for their abilities enabling the ice V to XIII phase transition.[65] HCl was reconfirmed to be the most effective dopant, which was attributed to a favorable combination of high solubility in ice V and its high acid strength that enables the formation of highly mobile $H_3O^+$ point defects. However, interestingly, lithium hydroxide doping was found to be of similar efficiency compared to doping with hydrofluoric acid. This makes ice XIII the first hydrogen-ordered phase of ice that can be prepared with both acid as well as base dopants.[65] The dramatically accelerated reorientation dynamics in HCl-doped ice V were confirmed using dielectric spectroscopy.[68]

HCl-doping also enabled the preparation of hydrogen-ordered ice XV from ice VI.[21] Consistent with an expansion of volume upon hydrogen ordering, the most ordered ice XV was obtained upon slow-cooling at ambient pressure.[21, 69] Yet, even this only leads to partially ordered ice XV according to neutron diffraction.[21, 69, 70] Calorimetric measurements have shown that slow-cooling HCl-doped $H_2O$ ice VI at ambient pressure leads to a ~50% loss of Pauling entropy.[65, 71] While base doping was found to be ineffective for obtaining ice XV, remarkably, doping with hydrobromic acid achieved a similar performance compared to hydrofluoric-acid doping.[65] It was speculated that the large bromide anions can potentially replace parts of the individual networks in ice VI (*cf.* Figure 2).

Why the various dopants display vastly different performances with respect to enabling hydrogen ordering of a given hydrogen-disordered phase of ice is still an open question. In ref. [65], three factors have been discussed that determine the performance of a dopant: (1) An effective dopant needs to soluble in the ice at high concentrations and (2) it needs to display a high acid or base strength in ice so that mobile $H_3O^+$ or $OH^-$ defects are created that accelerate the reorientation dynamics. (3) There may be a general difference with respect to the dynamics of the migration of $H_3O^+$ or $OH^-$ defects.

Regarding the question of acid strength, elegant fluorescence-quenching experiments were recently used to classify a wide range of weak and strong acids with respect to their acid strength in ice I*h*.[72, 73] The general trend is stronger acids in water also release more protons ($H^+$) into ice I*h*. Yet, the differences between strong and weak acids seem to be much less pronounced in ice compared to in water. Hydrofluoric acid seems to go slightly against this general trend and has been found to be a slightly stronger acid in ice compared to other weak acids. It would now be desirable to carry out similar experiments for the high-pressure phases as well.

On the basis of a new low-temperature endotherm, Gasser *et al.* recently claimed that cooling HCl-doped ice VI at pressures greater than 1.4 GPa leads to the formation of a new hydrogen-ordered phase of ice that is different from ice XV.[74] The new phase was suggested to be more thermodynamically stable than ice XV despite the fact that the new endotherm was found to be irreversible. Furthermore, it was noted that the new endotherm was only observed for HCl-doped $H_2O$ samples and not for the corresponding $D_2O$ samples, and this was attributed to a strong isotope effect.[74]

However, Rosu-Finsen *et al.* subsequently argued that the low-temperature endotherm is kinetic in origin and arises upon heating deep glassy states of hydrogen-disordered ice VI.[62] Consistent with the increase in volume upon hydrogen-ordering, the application of pressure suppresses the hydrogen ordering to ice XV, which then enables glassy states of ice VI to be reached at $T_g$(doped) along pathway (2) in Figure 4. The general effect of HCl-doping is a substantial decrease of the glass transition temperature from $T_g$(pure) to $T_g$(doped) estimated to be more than 30 K in case of ice VI.[62] Increasing the pressure enables more relaxed, deeper glassy states to be reached around $T_g$(doped). As encountered for a wide range of other glassy materials, deep glassy states display kinetic overshoot features upon heating connected with the underlying glass transition. Such endothermic features do not go along with an uptake of latent heat. It is therefore not possible to calculate entropy changes from such features and hence to correlate the peak area in a quantitative fashion with structural changes.

Consistent with this scenario, the low-temperature endotherms were found to either appear or disappear depending on the heating rate, and they could also be produced by prolonged annealing below $T_g$(doped) at ambient pressure. Furthermore, and in contrast to ref. [74], Rosu-Finsen *et al.* have shown that low-temperature endotherms can also be found for the corresponding $D_2O$ materials and neutron diffraction data of such a sample was shown to be consistent with deep glassy hydrogen-disordered ice VI.[62]

These recent findings illustrate that doping with HCl does not only enable hydrogen ordering of ices V, VI and XII. If the hydrogen ordering can be suppressed, for example by the application of pressure, then HCl-doping also allows accessing deep glassy states of a hydrogen-disordered ice at a significantly lowered glass transition temperature compared to the corresponding pure ice. Regarding the structure of deep glassy ice, it has been suggested that very local and spatially uncorrelated hydrogen ordering takes place so that the overall average structure is hydrogen-disordered.[62] If such a sample is subjected to subsequent hydrogen ordering, for example by annealing at lower pressures, most of the locally ordered domains would have to be undone in order to establish long-range hydrogen order.[62] Following the discovery of deep glassy states of ice VI, the question is now if such states can be realized for other ice polymorphs as well.

In addition to using acid and base dopants for achieving hydrogen ordering, it has recently been shown that adding polyethylene glycol to ice I*h* facilitates hydrogen-ordering to ice XI.[75] Emergent ferroelectric hydrogen ordering was recently found for thin ice I*h* films for temperatures up to 175 K, which seems to indicate that the presence of surfaces and interfaces can affect hydrogen-ordering processes as well.[76]

**VI. Solid-solutions of salts and the ice polymorphs**

Salts famously display very low solubilities in ice I*h* as observed during the freezing of sea water.[15] However, in stark contrast to ice I*h*, Klotz *et al.* have shown that ice VII can take up substantial amounts of lithium chloride and lithium bromide at ~1:6 LiX:$H_2O$ molar ratios.[77, 78] This was achieved by heating glassy aqueous solutions of these salts at high pressures. Compared to pure ice VII, 'salty' ice VII displays a significantly larger unit cell, considerable positional disorder of the water molecules and plasticity, which means that the energy barrier for the rotation of water molecules is greatly reduced due to the salt incorporation. Furthermore, the hydrogen-ordering phase transition from ice VII to ice VIII was also found to be completely suppressed in salty ice VII, and the ice VII to ice X transition pressure was shown to increase with increasing amounts of LiCl.[79] Salty ice VII may exist in icy moons and their different physical properties such as ionic conductivity compared to pure ice VII may influence their geophysics.[77, 78]

Compared to the lithium halides, the solubility of NaCl in ice VII was found to be much smaller.[80] However, the presence of NaCl also increased the transition pressure from ice VII to X.[81] Room temperature compression of aqueous NaF, NaCl, NaBr and NaI solutions showed that both the transition pressure from the liquid to ice VI as well as from ice VI to ice VII are affected by the sodium halides.[82, 83] Small amounts of $MgCl_2$ also seem to be soluble in ice VII.[84]

A remarkable exception to the low solubilities of salts in ice I*h* is $NH_4F$, which is miscible with ice I*h* across the entire composition range.[15] This is due to the very similar hydrogen-bonding properties of $NH_4^+$, $F^-$ and $H_2O$. Very recently, Shephard *et al.* explored the impact of small amounts of $NH_4F$ on the phase diagram of ice.[23] Remarkably, ice II was found to disappear in a selective fashion above 0.5 mol% $NH_4F$ to be replaced by either ice III or V, which readily form solid solutions with $NH_4F$. As shown schematically in Figure 5, the effect of incorporating ammonium and fluoride ions into a hydrogen-ordered ice structure, such as ice II, is that it creates hydrogen disorder. Within the defect-free and hydrogen-ordered ice II, the ions act as so-called 'topological charges' that disrupt the orientational order of the water molecules over long distances. This results in a substantial free-energy increase of the ice II, which explains its absence from the phase diagram in the presence of $NH_4F$. The $NH_4F$-doping induced disappearance of ice II from the phase diagram therefore exposes its strict topologically-constrained nature and the extremely long-range structural correlations of the ice II hydrogen-bond network.[23] In analogy to related magnetic spin-ice systems, it was suggested that the presence and nature of ice II may provide an explanation for the widely documented anomalies in the ice II pressure range, including those of liquid water (*cf.* Figure 2).[23] Furthermore, the impurity-induced disappearance of ice II now raises the prospect that specific dopants may not only be able to suppress certain phases but also induce the formation of new phases of ice in future studies.

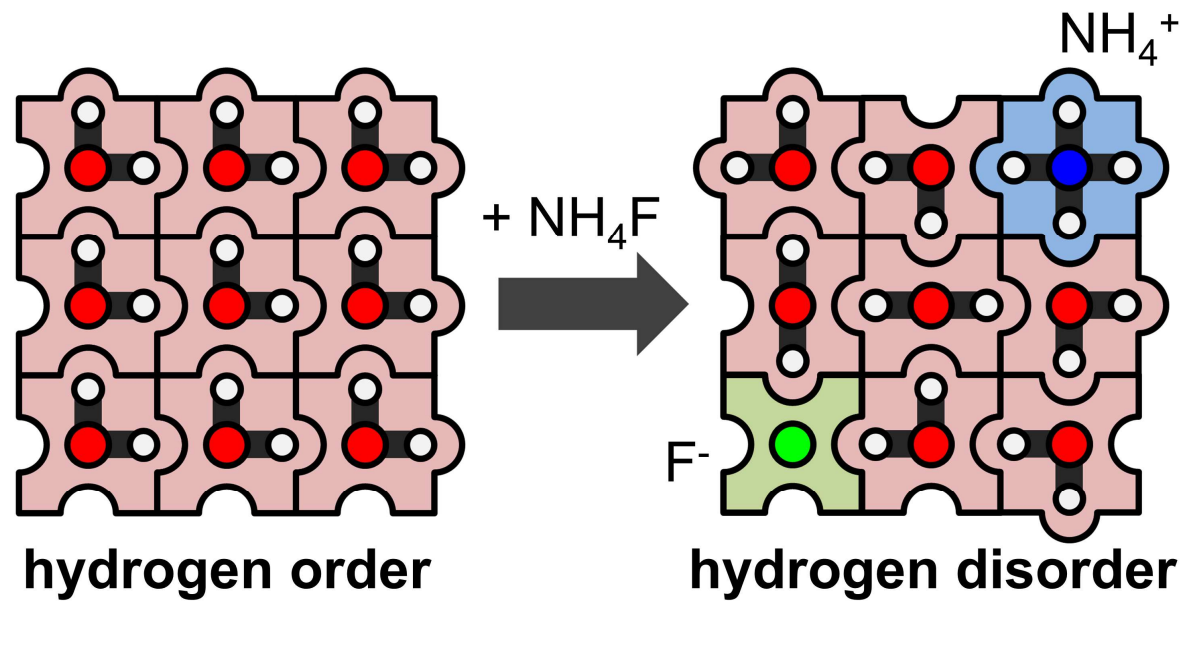

***Fig. 5*** *Schematic illustration of the hydrogen-disordering effect of $NH_4F$ substitution into hydrogen-ordered square ice. The circles indicate oxygen (red), hydrogen (white), fluorine (green) and nitrogen (blue) atoms. Adapted from ref. [23].*

## VII. Conclusions

As outlined in this perspective article, the exploration of 'chemical' dimensions of water's phase diagram has received a great deal of attention in recent years. This includes the emptying of clathrate hydrates to give new low-density ice polymorphs, the effects of acid / base dopants on hydrogen ordering processes as well as the solubility of ionic species in ice under pressure and the effects this has on the overall appearance of the phase diagram. Furthermore, high-pressure ice phases have recently been prepared within confinements including nanopores[85] and as micrometer-thick films sandwiched between sapphire discs.[86] Remarkably, the thin films of ices II, V and XIII displayed sufficient optical transmission properties so that 2D IR spectra could be collected.[86]

In addition to 'chemical' avenues, the exploration of extreme physical states will of course also continue to be a focus. This will certainly include pushing the boundaries with respect to achieving high pressures and temperatures, but also further studies of the phase diagram at negative pressures.[87] Most recently, the phase diagram of ice was explored up to 5000 K and 190 GPa using shock compression, which resulted in the discovery of superionic ice.[88] Using *in-situ* neutron diffraction in combination with a supported diamond-anvil cell, 52 GPa have recently been reached.[89] On the basis of the diffraction data, it has been suggested that interstitial hydrogen atoms exist in ice VII above 13 GPa. However, recent X-ray Raman measurements have challenged this finding.[90] Using *in-situ* dielectric dielectric[91] and Raman spectroscopy,[92, 93] defect and structural anomalies of ice VII have recently been detected at around or slightly above 10 GPa. A variety of phases have been predicted computationally as the terapascal regime is approached.[94-96] Experimentally, it has been shown that ice X persists up to 210 GPa,[26] and there is currently no evidence for any post-ice X phases of ice. In any case, despite almost 120 years of active research and a wide range of exciting new findings in recent years, it is clear that water's phase diagram still harbors many more secrets awaiting to be discovered.

## Acknowledgements

Funding is acknowledged from the Royal Society for a University Research Fellowship (UF100144), the Leverhulme Trust (RPG-2014-04) and the European Research Council under the European Union's Horizon 2020 research and innovation programme (grant agreement No 725271). Furthermore, I am grateful to J. J. Shephard, A. Rosu-Finsen, B. J. Murray, S. T. Bramwell, B. Slater, A. Michaelides, C. Andreani, R. Senesi, P. Hamm and G. P. Johari for helpful discussions.